\documentclass[
    ,final            
  ]
  {aipproc}

\layoutstyle{6x9}

\newcommand{\degr}{\mbox{$^\circ$}}

\newcommand{\po}{\mbox{$\pi^0$}}

\newcommand{\heeg} {\mbox{$\et\rightarrow e^+e^-\gamma$}}
\newcommand{\hmmg} {\mbox{$\et\rightarrow \mu^+\mu^-\gamma$}}
\newcommand{\heeee} {\mbox{$\et\rightarrow e^+e^-e^+e^-$}}
\newcommand{\heemm} {\mbox{$\et\rightarrow e^+e^-\mu^+\mu^-$}}
\newcommand{\hmmmm} {\mbox{$\et\rightarrow \mu^+\mu^-\mu^+\mu^-$}}
\newcommand{\hppmm} {\mbox{$\et\rightarrow \pi^+\pi^-\mu^+\mu^-$}}
\newcommand{\hee} {\mbox{$\et\rightarrow e^+e^-$}}
\newcommand{\hmm} {\mbox{$\et\rightarrow \mu^+\mu^-$}}
\newcommand{\ee} {\mbox{$e^+e^-$}}

\newcommand{\hppee}{\mbox{$\et\rightarrow\pi^+\pi^-e^+e^-$}}
\newcommand{\hppg}{\mbox{$\et\rightarrow \pi^+\pi^-\gamma$}}

\newcommand{\hppp}{\mbox{$\et\rightarrow \pi^+\pi^-\pi^0$}}
\newcommand{\hpppo}{\mbox{$\et\rightarrow \pi^0\pi^0\pi^0$}}
\newcommand{\hpppod}{\mbox{$\et\rightarrow \pi^0\pi^0\pi^0_D$}}

\newcommand{\eeg}{\mbox{$e^+e^-\gamma$}}

\newcommand{\et}{\mbox{$\eta$}}
\newcommand{\pdHeh}{\mbox{$pd\to^3$He\,$\eta$}}

\newcommand{\mum}{\mbox{$\mu^-$}}
\newcommand{\mup}{\mbox{$\mu^+$}}

\newcommand{\E}[1]{\mbox{$\times$10$^{#1}$}}

\begin{document}

\title{Leptonic decays of  the $\eta$ meson with the WASA detector at CELSIUS }


\author{M.~Berlowski}{
  address={The Andrzej Soltan Institute for Nuclear Studies, Warsaw, Poland}
}

\author{H.~Calen}{
  address={Uppsala University, Uppsala, Sweden}
}

\author{K.~Fransson}{
  address={Uppsala University, Uppsala, Sweden}
}

\author{M.~Jacewicz}{
  address={Uppsala University, Uppsala, Sweden}
}

\author{A.~Kupsc}{
  address={Uppsala University, Uppsala, Sweden}
}

\author{J.~Stepaniak$^*$\\ {\it for CELSIUS/WASA Collaboration}\\ }
{
  address={The Andrzej Soltan Institute for Nuclear Studies, Warsaw, Poland}
}
\begin{abstract}
Decay channels of the $\eta$ meson with at least one lepton pair in
the final state are discussed. Preliminary results on
electron-positron pair production from the \pdHeh~reaction from the
WASA experiment at CELSIUS are presented.
\end{abstract}

\maketitle


\section{Introduction}
The $\eta$ decay channels with lepton pairs are closely related to the
channels  with  real  photons.   The  direct  consequence  of  Quantum
Electrodynamics  is that  the process  with  a real  photon should  be
accompanied  by a process  where a  virtual photon  convert internally
into a  lepton pair.   This fact  was first pointed  out by  Dalitz in
1951~\cite{Dalitz:1951aj}  and therefore  the decays  with one  or two
lepton-antilepton pairs  are known as single or  double Dalitz decays.
Dalitz  decays can be  related to  the corresponding  radiative decays
using Quantum  Electrodynamics and introducing a function  of the four
momentum   transfer   squared   of   the   virtual   photon   ($q^2$):
$F(q_1^2,q_2^2)$ -- the  {\em Form Factor}.  In case  of Dalitz decays
$q^2$ is  equal to  the invariant  mass squared of  a lepton  pair and
$q^2\ge  4m_l^2$.  The  Form  Factor describes  the  structure of  the
transition  region and  it is  also used  to describe  processes where
$q^2<0$  (space-like virtual  photons). The  orientation of  the plane
defined by  the lepton momenta  is connected with the  polarization of
the virtual photon.
\begin{figure}[htbp]
a)\includegraphics[clip,width=0.24\textwidth]{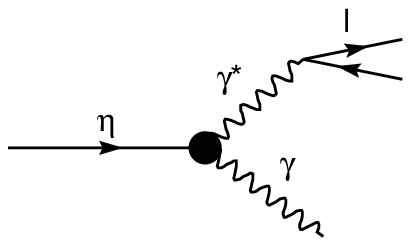}
b)\includegraphics[clip,width=0.24\textwidth]{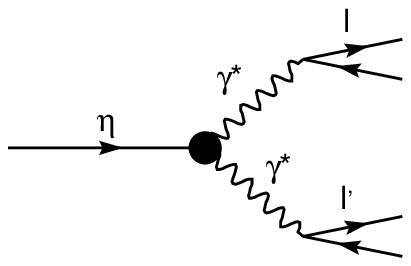}
c)\includegraphics[clip,width=0.24\textwidth]{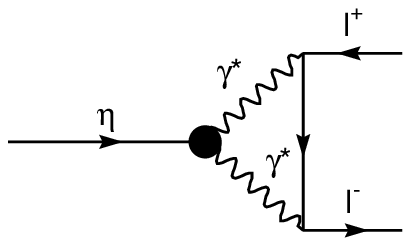}
\caption{\label{fig:hee2b} Diagrams for a) single and b) double Dalitz
decays of the pseudoscalar meson \et.  Diagram c) is for the
dominating conventional mechanism for \et~ meson decay into
lepton-antilepton pair.}
\end{figure}

The diagrams corresponding to the  conversion decays of \et~ are shown
in fig.~\ref{fig:hee2b}ab.  The diagram for the very rare decay into a
lepton-antilepton    pair   $P\rightarrow   l^+l^-$,    presented   in
fig.~\ref{fig:hee2b}c,  is forbidden  to proceed  via a  single photon
intermediate state.  The dominant  mechanism within the Standard Model
is the  second order electromagnetic process with  two virtual photons
$P\rightarrow\gamma^*\gamma^*$. In addition the decay is suppressed by
helicity  conservation.   Therefore  it  is  a  potentially  important
channel to look for effects of physics beyond the Standard Model.  The
imaginary part  of the decay amplitude  can be exactly  related to the
decay     $\et\to\gamma\gamma$.      The     measured     value     of
$\Gamma(\et\to\gamma\gamma)$ gives a lower limit (the unitarity limit)
of  BR($\eta   \to  e^+  e^-$)$\geq$1.8\E{-9}  \cite{Bergstrom:1982zq,
Landsberg:1986fd} which is a much lower value than for other decays of
\po\ and \et\ into lepton-antilepton  pairs.  This makes the $\eta \to
e^+  e^-$  decay  sensitive  to  contributions  from  non-conventional
effects  that would  lead to  a higher  branching ratio.   The current
experimental upper limit of $7.7 \times 10^{-5}$ (90\% CL) obtained by
the  CLEO II  collaboration  \cite{Browder:1997eu} is  four orders  of
magnitude higher.  The real part of the amplitude of the $\eta \to e^+
e^-$   decay  can   be   estimated  using   the   measured  value   of
BR($\et\to\mup\mum$)       \cite{GomezDumm:1998gw,      Savage:1992ac,
Ametller:1993we}  and gives a  prediction for  the branching  ratio of
about $6\times 10^{-9}$.  The estimated  BR is based on the assumption
that the  ratio between Im and Re  part of the decay  amplitude is the
same for $e^+e^-$~ and $\mup\mum$.

Dalitz decays of the \et\ allow studies of the structure of the
$\eta\rightarrow\gamma^*\gamma$ and $\eta\rightarrow\gamma^*\gamma^*$
vertex. An interesting question is how the Form Factor
$F(q_1^2,q_2^2)$ behaves as a function of the invariant mass squared
of the pairs $q^2$. Different forms of this dependence have been
proposed, the most common is predicted by the vector meson dominance
model.  Experimental information for the \et~ Dalitz decays are
scarce. The branching ratio for \heeg~ decay is known with rather
large error BR=$(6.0 \pm 0.8)\E{-3}$ \cite{Yao:2006px}.  The rate of
the double lepton pair decays of the \et\ has not been determined
experimentally, but different theoretical predictions can be found in
\cite{Jarlskog:1967, Picciotto:1993aa, Bijnens:1999jp,Borasoy:2007dw}.
The measured and calculated branching ratios for various lepton decay
channels are summarised in table \ref{tab:1}.
\begin{table}[htpb]
   \begin{tabular}{|l|l|l|}
    \hline
     Decay mode & Branching Ratio exp.    & Branching Ratio theor.\\
    \hline
    1. \heeg    &  $(6.0\pm0.8)\times10^{-3}$  &  $(6.37-6.57)\times10^{-3}$ \\
    2. \hmmg    &  $(3.1\pm0.4)\times10^{-4}$  &  $(2.1-3.05)\times10^{-4}$  \\
    3. \heeee   &  $<6.9\times10^{-5}$         &  $(2.52-2.64)\times10^{-5}$ \\
    4. \heemm   &  -       &  $(1.57-2.21)\times10^{-7}$       \\
    5. \hmmmm	&  -	   &  $2.4\times10^{-9}$  \\
    6. \hee     &  $<7.7\times10^{-5}$         & $1.8\times10^{-9}$~\tablenote{unitary bound}     \\
    7. \hmm     &  $(5.8\pm0.8)\times10^{-6}$  &  $4.4\times10^{-6}~^*$  \\
    8. \hppee   &  $(4.3\pm1.3\pm0.4)\times10^{-4}$  &  $3.1\times10^{-4}$  \\
    9. \hppmm	&  -	   &  $7.5\times10^{-9}$  \\

    \hline
   \end{tabular}
   \caption{\label{tab:1}The measured and calculated Branching Ratios for different \et~ decay channels with lepton pair(s).}
\end{table}      
\section{The Experiment}  
\begin{figure}[htbp]
\centering
\includegraphics[width=\textwidth]{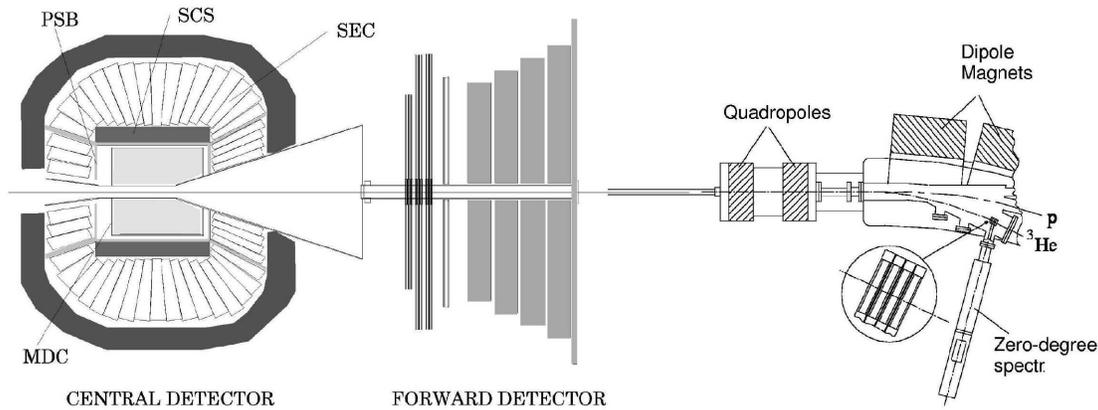}
\caption{ \label{fig:ts_bw2}
The WASA detector.}
\end{figure}
The experiment was  performed at the CELSIUS storage  ring in Uppsala,
using   the   WASA       detector      setup      (fig.~\ref{fig:ts_bw2})
\cite{Zabierowski:2002ah}.   Beam protons with a kinetic energy of  893 MeV
interacted  with frozen  droplets of  deuterium \cite{Ekstrom:2002ai}.
The \et~  mesons were  produced in the  reaction \pdHeh\ close  to the
\et~  production  threshold.   The  detection  of  $^3$He  ions  in  a
zero-degree   spectrometer  (tagging   detector)   provided  a   clean
\et~trigger  independent  of  decay  channel.  The  $^3$He  ions  were
identified  and their  energy  was measured  which  allowed to  select
cleanly  the \pdHeh\  reaction with  a background  of about  1\%.  The
tagging detector provided  a few triggers per second  (at a luminosity
of $ 5\times 10^{30}cm^{-2}s^{-1}$),  yielding on average one recorded
\et~  event   per  second.   During   two  weeks  of   the  experiment
(distributed  over half  a  year period)  nearly  $3\times 10^5$  \et~
events were collected.

The charged \et~  decay products are tracked using  a cylindrical mini
drift chamber (MDC),  consisting of 17 layers of  thin mylar tubes and
built around  a beryllium beam pipe  with 1.2 mm  wall thickness.  The
MDC is  placed inside of  a superconducting solenoid which  provides a
magnetic field  of 1T.  The MDC  is surrounded by a  barrel of plastic
scintillators.   An  electromagnetic  calorimeter consisting  of  1012
CsI(Na)  crystals   measures  the  energies  of   photons  and  impact
points.  Correlation   between  momentum,  measured   from  the  track
curvature  in  the  MDC  and  energy deposit  in  the  electromagnetic
calorimeter permits to distinguish electrons from charged pions if the
transverse particle momenta are larger than about 20 MeV/c.

In the  offline analysis,  events with at  least two  charged particle
tracks  reconstructed in  the MDC  were required  to originate  from a
point close  to the beam  target intersection region. Hit  clusters in
the calorimeter, without associated tracks  in the MDC and with energy
deposit larger  than 20  MeV were assumed  to originate  from photons.
Only  events  containing   decay  particle  candidates  with  balanced
electric charge were accepted for further analysis. The results on the
\hppee~     decay     channel     were    already     presented     in
ref.~\cite{Bargholtz:2006gz}.   Events with  a pair  of  charged decay
products with  opposite electric charges  can be attributed  either to
the  decay channels  with  two  charged leptons  or  to more  frequent
channels with two charged pions (\hppg~ and \hppp).

The following variables were used in the further data analysis:
\begin{enumerate}
\item   The  invariant  mass   of  the   pair  of   charged  particles
$M_{ee}$.  The electron masses  were assumed  in the  calculations.  A
clear peak  at the smallest possible  mass value is  expected for \ee~
Dalitz  pairs and  from conversion  of  real photons  in the  detector
material.
\item The invariant mass for all decay products. It was required to be
in the region of the \et~ mass.
\item The missing mass of the total decay product system ($MM_{\et}$).
It should be, within errors, equal to the mass of the $^3$He nucleus.
\item The ratio between the momentum measured in the drift chamber and
the energy of the shower in the electromagnetic calorimeter associated
with the  charged track $R_{p/E}$.  It permits  to distinguish between
electrons   and  charged   pions  for   the  particles   reaching  the
electromagnetic calorimeter.
\end{enumerate}
\section{Normalization}

\begin{figure}[htbp]
a)  \includegraphics[height=.3\textheight]{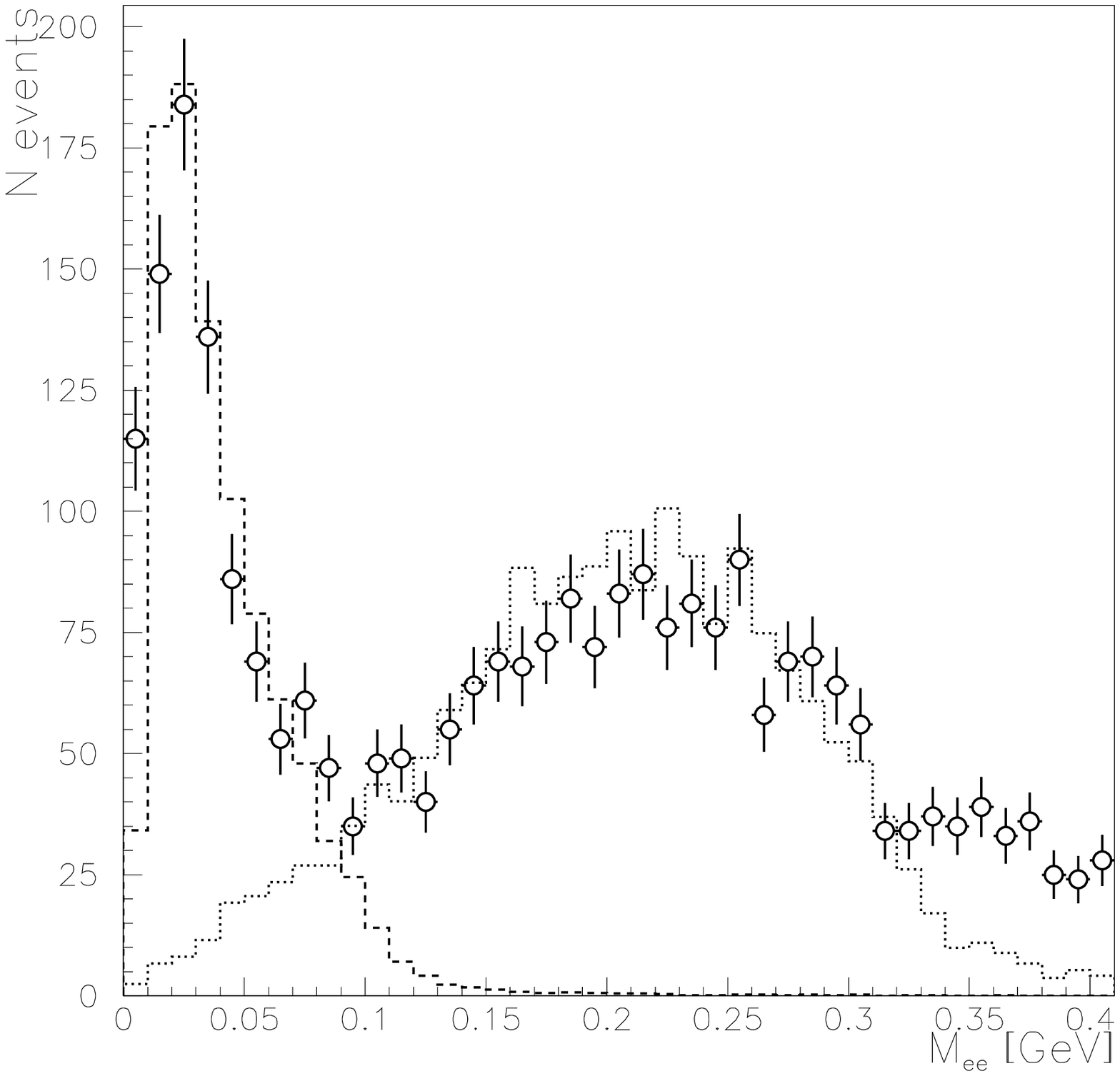}
b)  \includegraphics[height=.3\textheight]{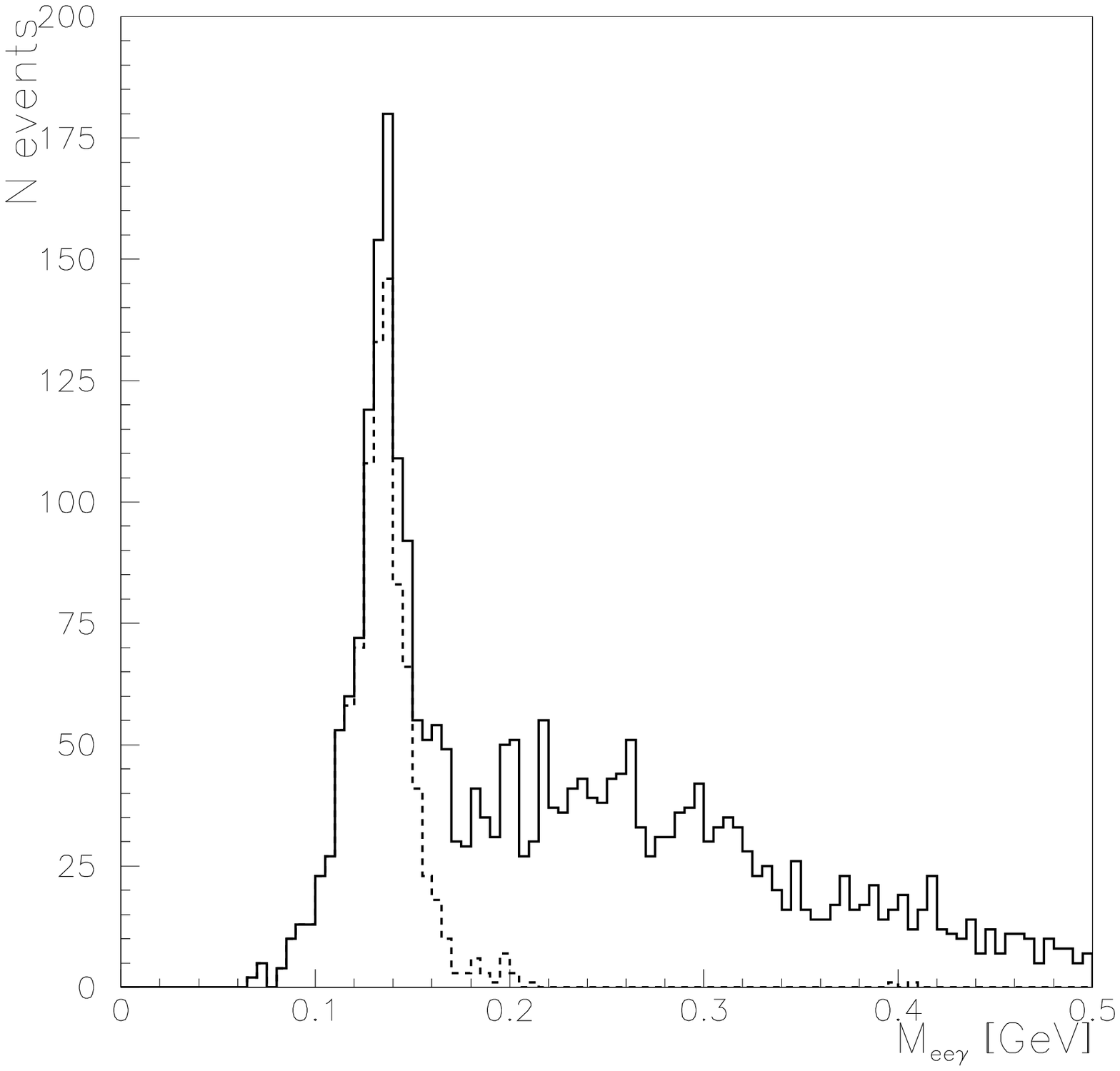}
\caption{ \label{fig:mee_3pi0} Event sample with two
tracks in MDC and at least four neutral clusters -- candidates for
\hpppod\ decay. Distributions of:  a) $M_{ee}$ -- experimental 
data (points), MC for \hpppod~
decay (dashed line), MC for decays involving a misidentified 
$\pi^+\pi^-$ pair (dotted
histogram). b) $M_{ee\gamma}$ --  all data (solid line) and  events with
$M_{ee}<0.1$~GeV/c$^2$ (dashed line).}
\end{figure}

In order  to normalize the branching  ratios of the  \et~ meson decays
involving an electron-positron pair  a monitoring process is needed to
check the  reconstruction efficiency for the  electrons.  Dalitz decay
of one of the three \po\  from the \hpppo\ decays provides an abundant
data set of events with  five photons and an electron-positron pair --
\hpppod.   The Dalitz  decay of  the \po\  meson has  been  studied in
detail both theoretically and  experimentally.  According to the Monte
Carlo  (MC) simulations  already the  class of  events with  more than
three  neutral  clusters is  dominated  by  this  decay channel.   The
invariant  mass  of the  two  charged  particles  for such  events  is
presented in fig.~\ref{fig:mee_3pi0}a.  The  peak at low masses can be
attributed to the \ee~ pairs from the neutral pion decay. The expected
distribution from  MC is  shown in the  same figure with  dashed line.
The  second maximum  is due  to the  decay channels  with $\pi^+\pi^-$
pair,  mainly \hppp\  (the corresponding  histogram in  the  figure is
normalized to the second enhancement).
Such interpretation is confirmed by the reconstructed invariant mass
of the \eeg\ system ($M_{ee\gamma}$) where the photon leading to the
mass value closest to the \po~mass is selected
(fig.~\ref{fig:mee_3pi0}b).  The distribution leads to a peak at
\po~mass when the $M_{ee}<0.1$~GeV/c$^2$ condition is applied.

\section{Results}

\begin{figure}[htbp]
  \includegraphics[clip,width=.8\textwidth,height=.3\textheight]{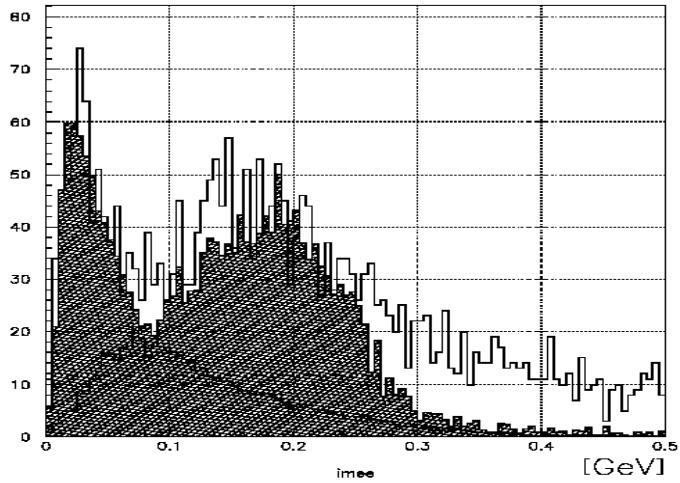}
\caption{ \label{fig:mee_eeg} $M_{ee}$ distribution for events with
less than four neutral clusters -- the initial selection for
\heeg\ decay. Histogram -- experimental data,
filled area -- MC simulation for \heeg~ decay and background from
decays involving a misidentified $\pi^+\pi^-$ pair. }
\end{figure}

The candidates  for \heeg\ decay are  searched in the  class of events
with less than four neutral  clusters. One of the clusters should have
energy deposit greater than 200  MeV. More than one cluster is allowed
in order not  to limit the acceptance for the  \heeg\ decay, where due
to electron or photon interaction in the calorimeter an additional hit
cluster  can be created.   In fig.~\ref{fig:mee_eeg}  the experimental
$M_{ee}$ distribution is shown  together with MC distributions for the
\heeg~ decay  and a  background from \et\  decays with  a $\pi^+\pi^-$
pair misinterpreted  as electrons.  The plot  is before identification
using conditions on the $R_{p/E}$ variable.

\begin{figure}[htbp]
a)\includegraphics[width=0.3\textwidth]{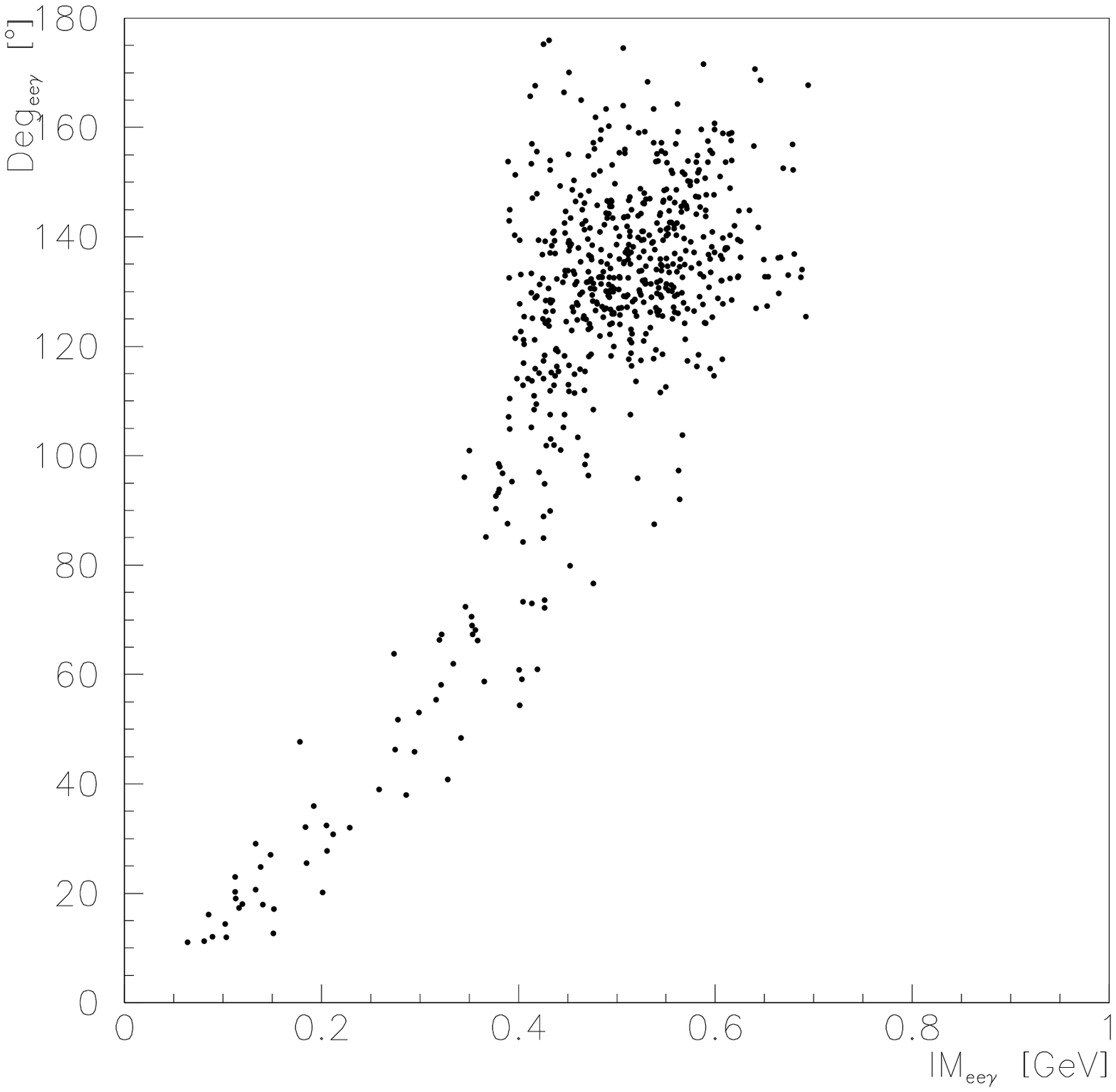}
b)\includegraphics[width=0.3\textwidth]{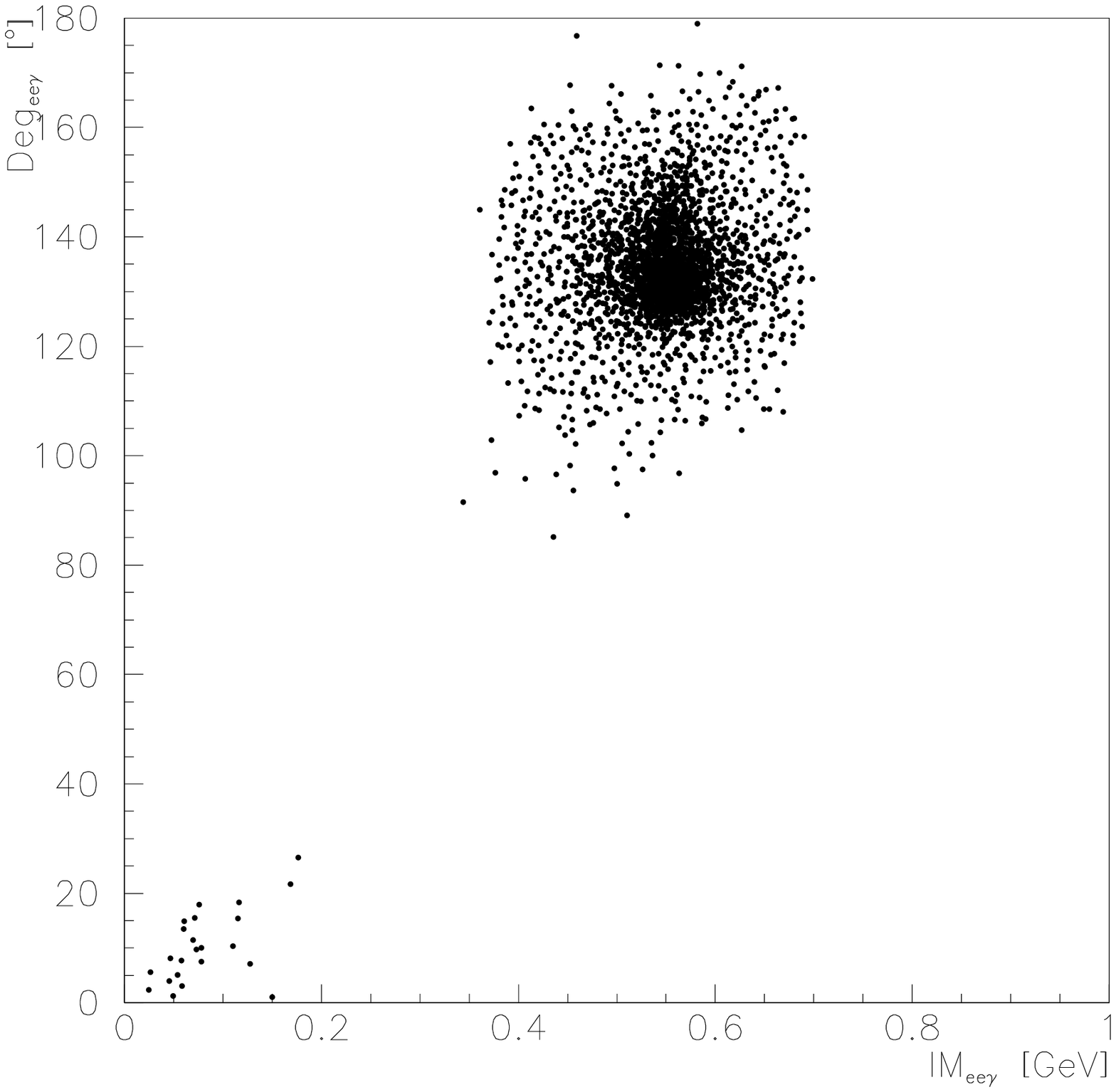}
\caption{\label{fig:e3}
Selected data  sample of the \heeg~ decay candidates:
relative angle between momentum of the 
electron-positron pair and the photon momentum vs 
$M_{ee\gamma}$ invariant mass, (a) data, (b) MC simulations for \heeg.}
\end{figure}

\begin{figure}[htbp]
\includegraphics[height=0.3\textwidth,width=0.6\textwidth]{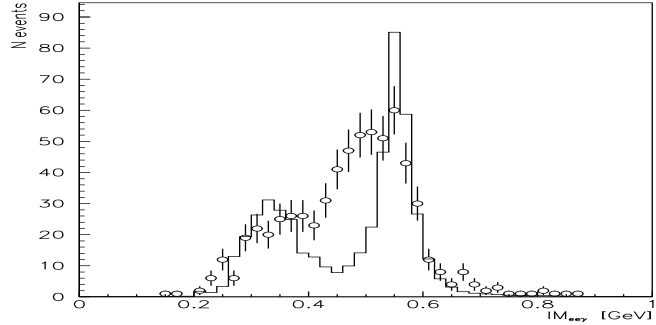}
\caption{\label{fig:e2} Selection of  the \heeg~ decay
candidates. The invariant mass $M_{ee\gamma}$ distribution:
experimental data (points); sum of the MC simulation of the \heeg\ and
the background channels (line).}
\end{figure}

In the final selection of  the \heeg~ decay candidates it was required
additionally that at least  one electron is identified using $R_{p/E}$
ratio and $M_{ee}<0.125$ GeV/c$^2$.  In fig.~\ref{fig:e3} the relative angle
between  the momentum  of the  electron-positron pair  and  the photon
momentum is plotted with  respect to the invariant mass $M_{ee\gamma}$
after the selection of the \heeg~ candidates.  Fig.~\ref{fig:e2} shows
the experimental  $M_{ee\gamma}$ distribution (points)  and the result
of the  MC simulation  of the signal  and the background  (solid line)
after the  selection cuts.  The  origin of the discrepancy  between MC
and the experimental data in the figure around 0.42 GeV/c$^2$ is under
study.  It can be due to the  interaction of the beam with rest gas in
the beam tube.

\begin{figure}[htbp]
\begin{minipage}[b]{0.5\textwidth}
\includegraphics[width=0.8\textwidth]{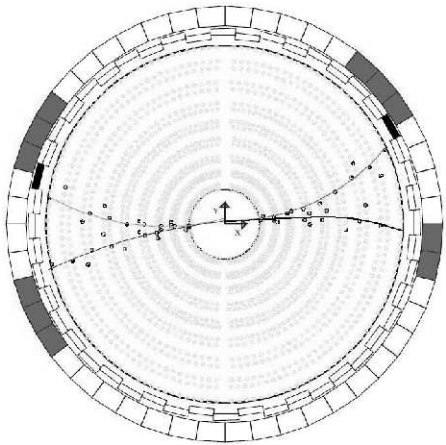}
\end{minipage}
\begin{minipage}[b]{0.5\textwidth}
\includegraphics[clip,width=\textwidth]{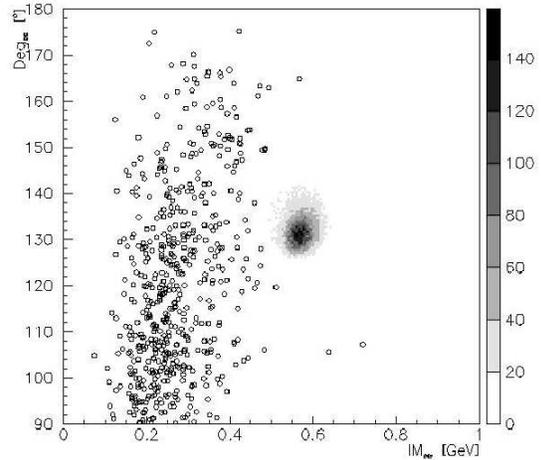}
\end{minipage}
\caption{\label{fig:event} (Left) Event display for an \heeee~
candidate event.  The shaded area in the outermost ring represents the
projection of the hit calorimeter crystals (the size of the crystals
and the radial position of the front faces is not to the scale).
 \label{fig:ee2} (Right)
Relative angle between electron and positron tracks vs
$M_{ee}$ for \hee~event sample selection: 
data -- scatter plot; MC simulation of the  \hee~decay  -- shaded area.}
\end{figure}

In a search for the \heeee\ decay, events with exactly four tracks in
the MDC and equal number of positive and negative charged particles
were selected.  According to the simulations 11\% of the reconstructed
\heeee\ events fulfill the following criteria:
\begin{enumerate}
\item the relative angle between electron and positron in both pairs
is less than 40\degr
\item the relative angle between the momenta of the two \ee\ pairs is
in the interval (110\degr-170\degr)
\item
the \et~ meson emission angle is smaller than 45\degr
\item the missing transverse momentum is less than 0.3 GeV/c
\item the missing mass $MM_{\et}$ is in the interval 2.65--2.90
GeV/c$^2$.
\end{enumerate}
In the data only two events passed all selection cuts. A schematic
view of the event display for one of the two candidates is shown in
fig.~\ref{fig:event}(Left).  The estimated background is 1.0 $\pm$ 0.3
events mainly from \heeg\ with a $\gamma$ conversion into $e^+e^-$ pair in
the detector material.

The \hee~ decay has a distinctive signature in the \pdHeh\ reaction
close to threshold: the emitted electron and positron have large
energies ($E>150$ MeV), are in plane with the beam and large
opening angle (about 134\degr).  In fig.~\ref{fig:ee2}(Right) the
$e^+e^-$ invariant mass is plotted as a function of the opening angle
of the electrons for the data sample and for the MC simulation of the
\hee~decay.  The selection included the electron identification
condition: $0.5< R_{p/E} <1.65$.  No events are observed in
\hee~signal region determined by MC simulations (corresponding to the
overall reconstruction efficiency of 41.2\%).

\section{Summary}
A data sample of about five hundred \heeg~ decays
events were collected.  This number is consistent with the
normalization based on \hpppod~ decay events. Two candidate events for
the double Dalitz decay mode \heeee\ were observed with an estimated
background of $1.0\pm0.3$ events. It is the first observation of this
\et~ decay channel. No candidates have been found for the \hee~ decay
in a sample of about 230k clean \et~events.
\begin{theacknowledgments}
We are grateful to the personnel at the The Svedberg Laboratory for their
support during the course of the experiment. The support from the Knut
and Alice Wallenberg Foundation (Sweden) and the Swedish Research
Council is acknowledged.
\end{theacknowledgments}



\bibliographystyle{aipproc}   

%
\bibliography{lit}
\end{document}